\documentclass[%
superscriptaddress,
reprint,
nofootinbib,
nobibnotes,
aps,
prl,
floatfix]{revtex4-1}
\usepackage{blindtext}
\usepackage{lmodern}
\usepackage{amsmath,bm}
\usepackage{amssymb,amsfonts}
\usepackage{xcolor}
\usepackage{graphicx}
\usepackage{dsfont}

\newcommand{\fjh}[1]{{ #1}}

\def\bea{\begin{eqnarray}}

\graphicspath{ {./images/} }

\begin{document}
\title{Magnetic Dipole Trapping Potential between Infinite Superconducting Plates}

\author{Francis J. Headley}
\email{francis.headley@uni-tuebingen.de}
\affiliation{Institut für Theoretische Physik, Eberhard-Karls-Universität Tübingen, 72076 Tübingen, Germany}



\date{\today}
\begin{abstract}
We derive the exact analytic form of the potential experienced by a magnetic dipole trapped between two infinite parallel superconducting plates using the method of image dipoles, providing a benchmark for numerical methods and a foundation for studying the stability and dynamics of magnetically levitated systems in precision measurements and fundamental physics experiments.
\end{abstract}
\maketitle
\section{Introduction}
The image charge method has long been used to solve boundary problems in electrostatics, particularly those concerning conductors and dielectrics \cite{Jackson1998}. More recently the image charge method has been used to approximate the potential of macroscopic magnets levitated in superconducting traps \cite{lin_theoretical_2006, vinante_levitated_2022}. These levitated systems boast high quality factors \cite{vinante_ultralow_2020,fuchs_magnetic_2023} and force sensitivity, being currently at the forefront of technological applications and the exploration of fundamental physics problems such as probing the quantum nature of weak forces such as gravity ~\cite{fuchs_magnetic_2023,prat-camps_ultrasensitive_2017, gonzalez-ballestero_levitodynamics_2021,bose_massive_2023,headley_quantum_2025} as well as its current use in the search for dark matter \cite{Amaral:2024rbj, carney2020}. 
For levitated systems in general, finding the potential for a trapped magnetic dipole is often done numerically, utilising finite element methods to solve the field equations, however in some special cases it is possible to obtain analytical solutions ~\cite{vinante_levitated_2022,lin_theoretical_2006}.
For macroscopic superconductors the London penetration depth can be assumed to be negligibly small and as such the superconductors can be assumed to be in an ideal Meissner state, a perfect diamagnet \cite{Tinkham1996}. With this assumption, it is possible to obtain an analytic form of the potential of a dipole levitated above an infinite superconducting  plate without ever having to solve the field equations \cite{lin_theoretical_2006,vinante_levitated_2022}. 

The purpose of this letter is to present an analytic form of the potential generated by a dipole trapped between two superconducting plates, which should serve as a good approximation of the trapping potential observed in real world experiments in fundamental physics, and furthermore provide a benchmark for numerical approximations of dipole potentials in more complicated geometries. We first present the derivation of the potential for a single dipole in the presence of one infinite superconducting plate. This solution is already present in the literature \cite{vinante_levitated_2022}, however the calculations serve as instruction for the derivation of the potential between two plates, which is presented directly after. These results are then compared with the Finite Element Method, which is a commonly used numerical technique for solving PDE's with different boundary conditions.

\section{Dipole above an infinite superconducting plane}

To begin, consider a magnetic dipole in the presence of an infinite superconducting plate. As the superconductor can be treated as a perfect diamagnet, the magnetic induction inside of the plate is zero. Thus, we only need to consider the magnetic field outside of the plate. There is no free electric current outside of the plate, therefore we can introduce the magnetic scalar potential $\Phi_\text{I}$ such that the total magnetic field is $\mathbf{B}=-\mu_0\nabla(\Phi_0+\Phi_\text{I})$ where $\Phi_0$ is the scalar potential of the magnetic dipole and $\Phi_\text{I}$ is the scalar potential generated by the induced currents on the surface of the superconductor. \fjh{Our dipole is situated near an infinite superconducting plate, defined at coordinate $z=a$. Noting that we have rotational symmetry of the geometry about the $z$‑axis, we work in cylindrical coordinates $(\rho,\phi,z)$, and assume that the dipole is oriented in the $(\rho,z)$-plane. In this case, we may set the angular component $\phi=0$, without loss of generality.} The scalar potential at a point $\boldsymbol{r}=(\rho,z)$ for a dipole $\boldsymbol{\mu}=\mu (\cos\beta_0,\sin\beta_0)$ located at $\mathbf{r}_0=(\rho_0,z_0)$ is given by 
\begin{equation}
\Phi_0(\rho,z)=\dfrac{\boldsymbol{\mu}\cdot(\mathbf{r}-\mathbf{r}_0)}{4\pi|\mathbf{r}-\mathbf{r}_0|^3}.
\end{equation}
 In order to calculate the induced scalar potential $\Phi_\text{I}$, we note that the Meissner effect implies the boundary condition on the induced magnetic field, that is $\mathbf{n}\cdot\mathbf{B}|_{z=a}=0$, where $\mathbf{n}$ is the unit vector normal to the surface of the superconductor, which in this case is a unit vector in the $z$ direction. To solve the boundary problem we place an image dipole located at $\mathbf{r}_1=(z_1,\rho_1)$
\begin{equation}
\Phi(\rho,z)=\Phi_0(\rho,z)+\Phi_\text{I}(\rho,z)=\dfrac{\boldsymbol{\mu}\cdot(\mathbf{r}-\mathbf{r}_0)}{4\pi|\mathbf{r}-\mathbf{r}_0|^3}+\dfrac{\boldsymbol{\mu}\cdot(\mathbf{r}-\mathbf{r}_1)}{4\pi|\mathbf{r}-\mathbf{r}_1|^3}.
\end{equation}
Where we write explicitly:
\begin{align}
\Phi_0(\rho,z)&=\dfrac{\mu(z-z_0)\sin\beta_0+\mu\fjh{\rho}\cos\beta_0}{4\pi(\fjh{\rho}^2+z^2+z_0^2-2zz_0)^{3/2}},\\ 
\Phi_\text{I}(\rho,z)&=\dfrac{\mu(z-z_1)\sin\beta_1+\mu\fjh{\rho}\cos\beta_1}{4\pi(\fjh{\rho}^2+z^2+z_1^2-2zz_1)^{3/2}}.
\end{align}
\fjh{We note here that both $\Phi_0$ and $\Phi_\text{I}$ satisfy the Laplace equation $\nabla^2\Phi=0$.} The correct boundary condition requires that the normal component of the magnetic field in the $z$ direction to be zero at the surface of the plate.
Taking the derivative with respect to $z$ and evaluating at $z=a$ we find the boundary term:
\begin{widetext}
\begin{align}
\partial_z\Phi|_{z=a}=&\dfrac{\mu\sin\beta_0}{4\pi(\fjh{\rho}^2+(z_0-a)^2)^{3/2}}-3(a-z_0)\left(\dfrac{\mu(a-z_0)\sin\beta_0+\mu\fjh{\rho}\cos\beta_0}{4\pi(\fjh{\rho}^2+(z_0-a)^2)^{5/2}}\right)\nonumber\\
&+\dfrac{\mu\sin\beta_1}{4\pi(\fjh{\rho}^2+(z_1-a)^2)^{3/2}}
-3(a-z_1)\left(\dfrac{\mu(a-z_1)\sin\beta_1+\mu\fjh{\rho}\cos\beta_1}{4\pi(\fjh{\rho}^2+(z_1-a)^2)^{5/2}}\right). \label{boundary1}
\end{align}
\end{widetext}
To set the term \eqref{boundary1} to zero, we fix the $z$-coordinate of the image charge to be $z_1=2a-z_0$, and its orientation $\beta_1=-\beta_0$. This is tantamount to placing an oppositely oriented dipole of equal strength equidistant on the other side of the plate. The scalar potential of the image charge is thus:
\begin{equation}
\Phi_\text{I}(\rho, z)=\dfrac{-\mu(z-(2a-z_0))\sin\beta_0+\mu\fjh{\rho}\cos\beta_0}{4\pi(\fjh{\rho}^2+(z-(2a-z_0))^2)^{3/2}}.
\end{equation}
\fjh{Because both $\Phi_0$ and $\Phi_{\text{I}}$ satisfy Laplace's equation and the combined potential satisfies the boundary condition at $z = a$, the uniqueness theorem for Laplace's equation guarantees that this solution is unique \cite{Jackson1998}. We now set $\rho_0=0$ without loss of generality.} The potential experienced by the dipole above the plate is calculated using \cite{lin_theoretical_2006}:
\begin{equation}
    U(z_0,\beta_0)=-\frac{1}{2}\boldsymbol{\mu}\cdot\mathbf{B}_\text{I},
\end{equation}
where $\mathbf{B}_\text{I}=-\mu_0\boldsymbol{\nabla}\Phi_\text{I}$. After some simple calculation, the potential experienced by the dipole levitated above an infinite superconducting plate may be expressed as:
\begin{equation}
U(z_0,\beta_0)=\dfrac{\mu_0\mu^2}{64\pi(\fjh{z_0}-a)^3}(1+\sin^2\beta_0). \label{oneplate}
\end{equation}
As we have translational invariance along the $\rho$ axis, we have set $\rho=0$. Although this result is known in the literature \cite{vinante_levitated_2022}, the derivation will serve as a basis for the calculations in the next section.

\section{Dipole trapped between two superconducting plates}\label{twoplates}
To calculate the potential between two plates located at $z=a$ and $z=b$, we generalize the method from the single-plate case by accounting for an infinite series of image dipoles. This iterative process ensures that the boundary conditions are consistently satisfied on both superconducting surfaces. To begin, we introduce two image dipoles with potentials $\Phi_1$ and $\Phi_{-1}$ at  $z_{1}=2a-z_0$ and $z_{-1}=2b-z_0$ respectively:
\begin{align}
\Phi_0=&\dfrac{\mu(z-z_0)\sin\beta_0+\mu\rho\cos\beta_0}{4\pi(\rho^2+z^2+z_0^2-2zz_0)^{3/2}},\\
\Phi_1=&\dfrac{-\mu(z-(2a-z_0))\sin\beta_1+\mu\rho\cos\beta_1}{4\pi(\rho^2+z^2+(2a-z_0)^2-2z(2a-z_0))^{3/2}},\\
\Phi_{-1}=&\dfrac{-\mu(z-(2b-z_0))\sin\beta_{-1}+\mu\rho\cos\beta_{-1}}{4\pi(\rho^2+z^2+(2b-z_0)^2-2z(2b-z_0))^{3/2}}.
\end{align}
 We introduce these dipoles with the same intention as in the previous section, to set the normal component of the $B$-field to zero at the plates $\partial_z\Phi|_{z=a/b}=0$, however this fails immediately. To see this, we sum these terms $\Phi=\Phi_0+\Phi_1+\Phi_{-1}$ and take the derivative with respect to $z$.

\begin{figure*}[t]
\centering
\includegraphics[width=0.8\textwidth]{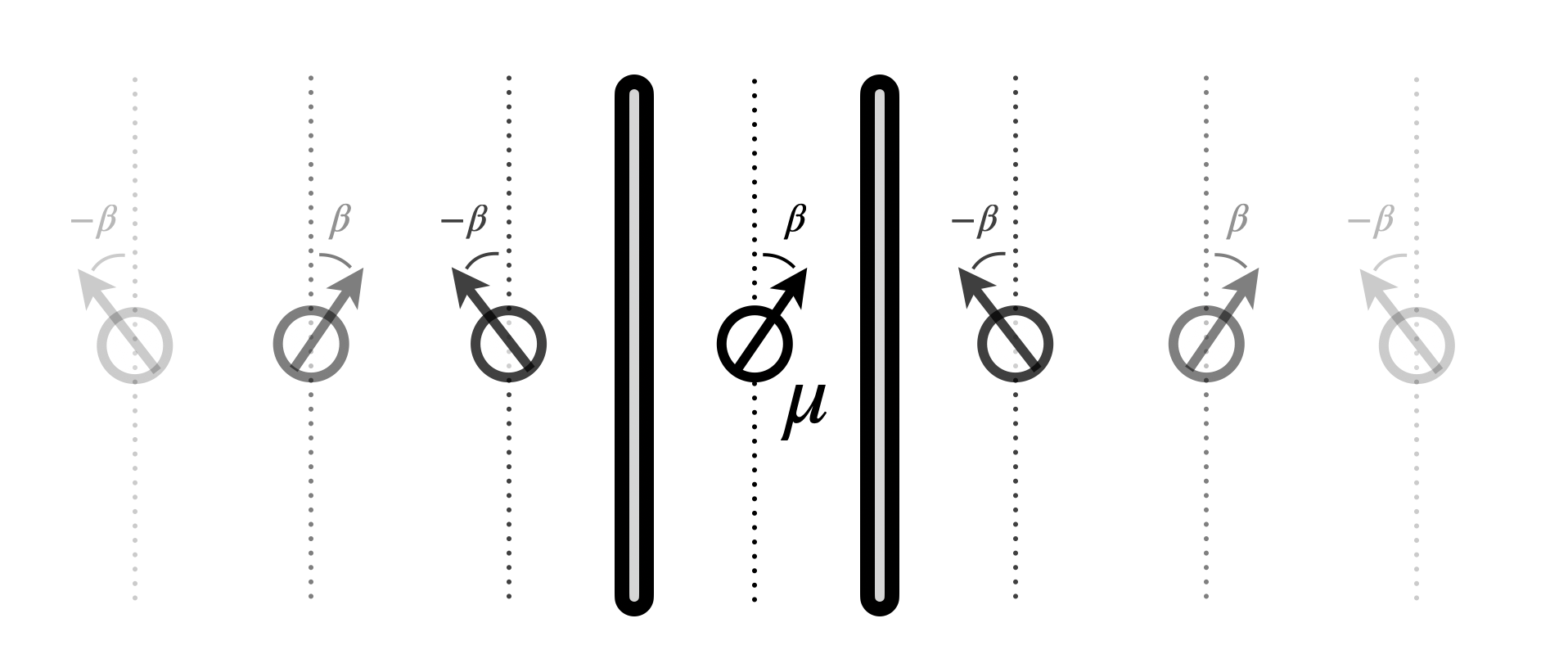}
\caption{Schematic representation of the infinite stack of image charges used to model the magnetic dipole $\mathbf{\mu}$ trapped between two infinite superconducting plates. The image dipole method is used to represent the influence of the superconductors, creating an infinite series of mirror dipoles that enforce the boundary conditions.  }\label{fig:diag}
\end{figure*}
Evaluating at the $a$-plate we find that the image dipole located at $z_{-1}$ contributes a non-zero term, similarly the image dipole located at $z_{1}$ contributes a non-zero term on the $b$-plate. Explicitly:
\begin{widetext}
\begin{align}
\partial_z\Phi|_{z=a}=&\dfrac{\mu\sin\beta_{-1}}{4\pi(\rho^2+(a-(2b-z_0))^2)^{3/2}}-3(a-(2b-z_0))\left(\dfrac{\mu(a-(2b-z_0))\sin\beta_{-1}+\mu\rho\cos\beta_{-1}}{4\pi(\rho^2+(a-(2b-z_0))^2)^{5/2}}\right), \label{t1}\\
\partial_z\Phi|_{z=b}=&\dfrac{\mu\sin\beta_1}{4\pi(\rho^2+(b-(2a-z_0))^2)^{3/2}}-3(b-(2a-z_0))\left(\dfrac{\mu(b-(2a-z_0))\sin\beta_1+\mu\rho\cos\beta_1}{4\pi(\rho^2+(b-(2a-z_0))^2)^{5/2}}\right).\label{t2}
\end{align}
\end{widetext}
To set these terms to zero we must modify our scalar potential to include two new image charges $\Phi\to\Phi'=\Phi+\Phi_{-2}+\Phi_2$. To cancel the term \eqref{t1} we add a dipole $\Phi_{-2}$ at $z_{-2}=2a-2b+z_0$ with orientation $\beta_{-2}=\beta_0$. Similarly to cancel the term \eqref{t2} we introduce a fourth dipole $\Phi_2$ at $z_2=2b-2a+z_0$ with orientation $\beta_2=\beta_0$. However, we are not out of the woods yet. Indeed, if we evaluate the derivative of our new scalar potential at both plates  $\partial_z\Phi'|_{z=a/b}$ we find that the image charges we introduced to generate counter terms, will themselves require counter terms. It is easy to see how the charges introduced to generate counter terms require counter terms themselves. Luckily for us, we can do this forever. Repeating the process with our new potential $\Phi'$  we see that we must introduce two new dipoles at $z_{-3}=4a-2b-z_0$ and $z_3=4b-2a-z_0$ with orientations $\beta_{-3}=\beta_3=-\beta_0$.  We repeat this process ad infinitum and enumerate it in a clever way. The 'counter-clockwise oriented' $\beta_0$ image dipoles $\Phi_{n+}$ and the 'clockwise-oriented' $-\beta_0$  image dipoles $\Phi_{n-}$ must be located at positions
\begin{align}
z_{n+}&=2n(b-a)+z_0,\nonumber \\ z_{n-}&=2n(b-a)+2b-z_0, \qquad n\in\mathbb{Z},
\end{align}
respectively, such that all terms vanish on the plates. We note that the dipole $z_{0+}$ is the original dipole located at $z_0$. Combining the above and summing over our infinite image charges, we arrive at the expression for the total scalar potential: 
\begin{align}
\Phi_\text{tot}(\rho,z)=\sum_{n\in\mathbb{Z}}&\left(\dfrac{\mu(z-z_{n+})\sin\beta_0+\mu\rho\cos\beta_0}{4\pi(\rho^2+z^2+z_{n+}^2-2zz_{n+})^{3/2}}\right. \nonumber\\
&\quad \left. +\dfrac{\mu(z_{n-}-z)\sin\beta_0+\mu\rho\cos\beta_0}{4\pi(\rho^2+z^2+z_{n-}^2-2zz_{n-})^{3/2}}\right).
\end{align}

We can decompose the total scalar potential $\Phi_\text{tot}=\Phi_0+\Phi_\text{I}$ where $\Phi_0$ is the contribution from the dipole and $\Phi_\text{I}= \sum_{n\neq0}\Phi_{n+}+\sum_n\Phi_{n-}$ is the contribution from the image charges. To calculate the potential experienced by the dipole, we calculate $U(z,\beta)=-\frac{1}{2}\boldsymbol{\mu}\cdot\boldsymbol{B}_\text{I}$, where $\boldsymbol{B}_\text{I}=-\mu_0\boldsymbol{\nabla}\cdot\Phi_\text{I}$ \cite{lin_theoretical_2006}. Evaluating at $z=z_0$ and setting $\rho=0$ we obtain the trapping potential of the dipole between two superconducting plates:

\begin{figure*}[t]
\includegraphics[width=0.9\columnwidth]{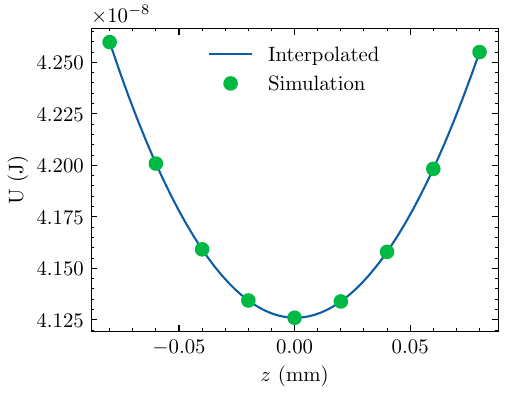}   \includegraphics[width=0.9\columnwidth]{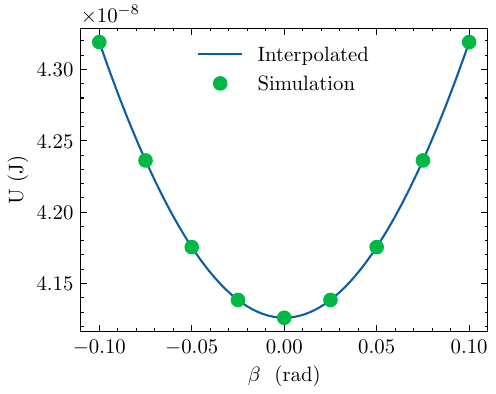} 
\caption{Finite element method (FEM) simulations of a magnetic dipole trapped in a hollow superconducting cylinder near equilibrium. The left and right plots correspond to the $z$ and $\beta$ coordinates, respectively. The data, obtained by numerically solving Maxwell’s equations using FEniCS, is fitted with quadratic functions to extract the trapping frequencies via Eq.~\eqref{frequ}.
}\label{fig:plot}
\end{figure*}

\begin{equation}
U
=\dfrac{\mu_0\mu^2}{8\pi}\left(\dfrac{\cos^2\beta_0-2\sin^2\beta_0}{4|b-a|^3}\zeta(3)+\sum_{n\in \mathbb{Z}}\dfrac{\sin^2\beta_0+1}{|z_0-z_{n-}|^3}\right), \label{pot}
\end{equation}
where $\zeta(n)$ is the Riemann zeta function. This is our main result. The expression \eqref{pot} is divergent only if $z_0=b\ \text{or}\ a$, and is convergent otherwise. Furthermore, we note that in the limit $b\to\infty$, the two plate potential \eqref{pot} converges to the single plate potential \eqref{oneplate}. \fjh{Assuming that the infinite plates are located the same distance from the origin, that is to say at $z=b$ and $z=-b$,
the resonance frequencies of the $z$ and $\beta$ modes may be calculated via the spring constants $\omega_z=\sqrt{k_z/m}$ and $\omega_\beta=\sqrt{k_\beta/I}$, where $I$ is the moment of inertia of the magnet, which for a solid sphere magnet of radius $r$ is $I=\frac{2}{5}mr^2$, $m$ is the mass of the magnet, and $k_z$, $k_\beta$ are the spring constants calculated via the second derivatives of \eqref{pot}:
\begin{equation}    k_z=\left.\dfrac{\partial^2U}{\partial z_0^2}\right|_{z_0=0,\beta_0=0}\qquad k_\beta=\left.\dfrac{\partial^2U}{\partial \beta_0^2}\right|_{z_0=0,\beta_0=0}.
\end{equation}
as in \cite{vinante_levitated_2022}}. Thus we find the resonance frequencies to be
\begin{equation}    \omega_z=\left.\sqrt{\dfrac{1}{m}\dfrac{\partial^2 U}{\partial z_0^2}}\right|_{z_0=0,\beta_0=0} \qquad \omega_\beta=\left.\sqrt{\dfrac{1}{I}\dfrac{\partial^2 U}{\partial \beta_0^2}}\right|_{z_0=0,\beta_0=0} \label{frequ}
\end{equation}
The $z_0$ and $\beta_0$ mode frequencies may be extracted:
\begin{equation}
      \omega_z=\sqrt{\dfrac{93}{256}\dfrac{\zeta(5)}{\pi}\dfrac{\mu_0\mu^2}{b^5m}},
      \qquad \omega_\beta=\sqrt{\dfrac{5}{64}\dfrac{\zeta(3)}{\pi}\dfrac{\mu_0\mu^2}{b^3mr^2}}. \label{freq}
\end{equation}
This formulation allows us to evaluate the magnetic potential at any point between the plates with arbitrary precision. Setting $z_0=0$ in \eqref{pot}, we find a simple form of the rotational potential:
\begin{equation}    U(\beta_0)=\dfrac{\mu_0\mu^2\fjh{\zeta(3)}}{128b^3\pi}(5-\cos(2\beta_0))
\end{equation}
which is a double well potential with minima located at $\beta_0=0, \ \pi$, which corresponds to both orientations of the magnet parallel to the plates.

\section{Numerics}
A useful point of comparison for Eq.\eqref{pot} and Eq.\eqref{freq} would be another commonly used method of extracting trapping frequencies for levitated particles. One example of such a method is Finite Element Method (FEM), which has been used to numerically approximate trapping frequencies for experiments involving levitated superconductors ~\cite{timberlake_acceleration_2019,vinante_ultralow_2020}. FEM was selected due to its robustness in solving partial differential equations with boundary conditions, as encountered in the case of dipoles near superconducting planes. By discretizing the spatial domain of the problem into finite chunks, FEM allows for precise calculation of magnetic potential. Treating the levitated particle as a point dipole and assuming that the London penetration depth is negligibly small, the system is modelled using the following differential equations:
\begin{align}
    \nabla\times \boldsymbol{B}_\text{I} = 0 \qquad \  &\forall x \in \Omega,\nonumber\\
    \boldsymbol{n}\cdot \boldsymbol{B}_\text{I}=-\boldsymbol{n}\cdot \boldsymbol{B}_0 \ \  &\forall x \in \partial \Omega, \label{maxwell}
\end{align}
where  $\mathbf{B}_0$ is the magnetic field generated by the dipole, $\mathbf{B}_\text{I}$ is the induced magnetic field, $\boldsymbol{n}$ represents the unit vector normal to the surface of the superconductor, $\partial\Omega$ represents the surface of the superconductor and $\Omega$ the space outside of the superconductor, in our case the interior of a hollow superconducting cylinder where the dipole is located. 

To solve this magnetostatics problem, we use FEniCS, a Python package for finite element method to solve PDE's ~\cite{AlnaesEtal2015}. The trapping frequencies are obtained by numerically solving Maxwell’s equations \eqref{maxwell} at points surrounding the dipole’s equilibrium position, for both the $z$ and $\beta$ coordinates. At each point, the potential energy is calculated and fitted to a quadratic function, allowing the trapping frequencies to be determined via Eq.~\eqref{frequ}. 
With a Neodymium magnet of density $\rho=7420\ \text{kg}\ \text{m}^{-3}$, $\mu=5.222\times10^{-5}\ \text{A}\ \text{m}^2$, radius $r\approx2.4\times10^{-4}\ \text{m}$, a mass of $m=4.3\times10^{-7}\ \text{kg}$, and assuming the distance from the centre of the trap to the superconductor is $b=1\ \text{mm}$, the magnetic field induced by a point-like dipole was simulated. Although it is not possible to simulate two infinite plates using FEniCS, an approximation was implemented by using a wide cylinder of height $2b=2\ \text{mm}$ and radius $25\ \text{mm}$. The results of this numerical procedure are plotted in Fig \ref{fig:plot}, and the frequencies extracted the interpolated data are displayed in Table \ref{tab}.
\begin{center}
\begin{table}[h]
\begin{tabular}{||c c c c c c c c c||} 
 \hline
&   && Analytic  &&  FEniCS  &&  $\text{Diff}_\text{rel}$&\\ [0.5ex] 
 \hline\hline
& $f_z$ && 155.545 && 154.673 && 0.56\% &\\ 
 \hline
& $f_\beta$ && 323.659 && 308.899 && 4.78\% &  \\ [1ex] 
 \hline
&  $U$ && $4.096\times10^{-8}$ && $4.126\times10^{-8}$ && 0.73\%  & \\ [1ex] 
 \hline
\end{tabular}
\caption{Results for the $z$ frequency, $\beta$ frequency, and potential energy, obtained using the analytic expression in Eq.~\eqref{pot} as well as the the finite element method (FEniCS).
}\label{tab}
\end{table}
\end{center}
As we can see, for the calculation of the trapping frequency of the $z$-mode $f_z=\omega_z/2\pi$, the analytic result deviates only by approximately $\sim0.5\%$ from the value calculated using FEM. For the $\beta$-mode trapping frequency the deviation from the FEM data is far larger, approximately $\sim5\%$. One possible source of this deviation can be attributed to the additional geometry used when calculating the trapping frequencies using FEniCS, which in our case was a flattened cylinder. The additional circular walls of the trap, which are not considered in the analytic results, also contribute to the induced magnetic field and thus the trapping frequencies. This additional contribution to the field is more present in the $\beta$-mode than in the $z$-mode. It is also seen that increasing the radius of the simulated superconducting region systematically reduces the discrepancy between the numerical and analytical results, as the finite-size effects diminish and the boundary conditions more closely approximate the idealized analytic geometry.

\section{Conclusion}
We have derived the analytic potential experienced by a magnetic dipole positioned between two infinite parallel superconducting plates. By extending the image method commonly used in electrostatics, we formulated a systematic approach to construct the infinite series of image dipoles required to satisfy the boundary conditions imposed by the superconducting plates. This approach allowed us to obtain an exact analytical form of the trapping potential. Our results provide a useful benchmark for numerical simulations of levitated magnetic dipoles in superconducting environments. The derived potential serves as a step toward improving the accuracy of models used in high-precision experiments involving magnetic levitation, in particular those investigating weak forces such as gravity, applications to dark matter searches, as well as modelling of mesoscopic quantum tunnelling between orientations of the levitated magnet. 

\textit{Acknowledgements --} 
We thank Alessio Belenchia, Fabian Müller, Hendrik Ulbricht, Tim Fuchs, Chris Timberlake, Elliot Simcox, and Denis Uitenbroek for fruitful discussions. We further thank  Daniel Braun, Alessio Belenchia and Sophie Heinrich for feedback on the draft of this paper, as well ast Giulio Gasbarri for help with the FEM code. We acknowledge the EU EIC Pathfinder project QuCoM (101046973).

\bibliographystyle{siam}
\bibliography{refs.bib}

\end{document}